# The relativistic quantum hydrodynamic representation of Klein-Gordon equation

Piero Chiarelli

*National Council of Research of Italy, Area of Pisa, 56124 Pisa, Moruzzi 1, Italy Interdepartmental Center "E.Piaggio" University of Pisa*

Phone: +39-050-315-2359
Fax: +39-050-315-2166

Email: pchiare@ifc.cnr.it.

**Abstract:** The quantum hydrodynamic-like equations for two real variables (i.e., the phase and the amplitude of the wave function) of the relativistic Klein-Gordon equation are derived in the present paper. The paper also shows that in classical limit the hydrodynamic Klein-Gordon equations lead to the Madelung pseudo-potential [1] as well as to the quantum pseudo potential for a charged particle given by Janossy [2] and by Bialynicki et al [3]. The origin of the non-local interactions of quantum mechanics in the hydrodynamic model is discussed both for free and charged particles.

# 1. Introduction

In the classical limit the Schrödinger equation is a differential equation where the non-local character of evolution is determined by the initial and boundary conditions that must be defined for describing a physical problem.

In the hydrodynamic quantum equations (HQEs) the non-local restrictions come by applying the quantization of vortices [3] and by the elastic-like energy arising by the quantum pseudo-potential but not from boundary conditions (that in the HQEs for instance are the vanishing of density $|\psi|$ at infinity and initial conditions).

In the case of charge particles the non-local properties for the Schrödinger equation come also from the presence of the electromagnetic (em) potentials that depend by the intensities of em fields in a non-local way (e.g., Aharonov –Bohm effect).

In the corresponding hydrodynamic equations the em potentials do not appear [3] but only in local way through the strength of the em fields. Being so, the hydrodynamic equations exhibit more clearly the origin of the non-local character of quantum physical law than in the Schrödinger equation.

Even if the hydrodynamic and the wave descriptions are perfectly equivalent only a crazy man would prefer to solve the non-linear HQEs [1-4] instead of the Schrödinger or Pauli ones.

The mathematically more clear statements of non-local restrictions of the HQEs and their classical-like structure makes it suitable for the achievement of the conceptual connection between quantum concepts (probabilities) and classical ones such as (e.g., trajectories) [5-6]. This fact makes the HQEs more suitable in describing phenomena at the edge between the quantum and classical mechanics such as the description of the dispersive effects in semiconductors [7,8] critical phenomena [9].

The advantage of HQEs in managing the non-local quantum character becomes more evident in system larger than a single atom when fluctuations becomes important [10] or when we want to investigate the effect of them onto the coherence of quantum non-local evolution [11], a field of great interest in the scientific community [12-17].

The stochastic generalization of the HQEs equations [11] can lead to a clearer insight about the interplay between fluctuations and the quantization conditions explaining why it survives on atomic scale (i.e., Compton length) while it is disrupted at large one in a fluctuating environment.

The availability of having the quantum hydrodynamic description of the Klein-Gordon, hence, one is not just a generic additional theoretical tool but it can be very useful in investigating the problem of quantum entanglement in a fluctuating environment in the relativistic limit (including photons) and the breaking of quantum non-locality on large scale.

## 2. The HQE from the Schrödinger equation

In this paragraph we define a general standard procedure to derive the hydrodynamic quantum equations from the quantum wave ones and then apply it to the Klein-Gordon (K-G) equation.
In this paragraph we analyze the procedure onto the Schrödinger equation.
The HQE-equations are based on the fact that the Schrödinger equation, applied to a wave function $\Psi_{(q,t)} = |\Psi| \exp[\frac{iS_{(q,t)}}{\hbar}]$, is equivalent to the motion of a fluid with particle density $n_{(q,t)} = |\Psi|^2$ and a velocity $\dot{q}_{(q,t)} = \frac{\nabla_q S_{(q,t)}}{m}$, governed by the equations [3,18]

$$\partial_t n_{(q,t)} + \nabla_q \bullet (n_{(q,t)} \nabla_q \dot{q}) = 0, \tag{1}$$

$$\dot{q} = \nabla_p H = \frac{\nabla S_{(q,t)}}{m}, \tag{2}$$

$$\dot{p} = -\nabla(H + V_{qu}), \tag{3}$$

with

$$\nabla_p \equiv (\frac{\partial}{\partial p_1}, ..., \frac{\partial}{\partial p_{3n}}) \tag{4}$$

where

$$H = \frac{p \bullet p}{2m} + V_{(q)} \tag{5}$$

is the Hamiltonian and $V_{qu}$ is the quantum pseudo-potential that reads

$$V_{qu} = -(\frac{\hbar^2}{2m}) n^{-1/2} \nabla \bullet \nabla n^{1/2}. \tag{6}$$

Equation (1-3) with the identities

$$p = m\dot{q} = \nabla S \tag{7}$$

where

$$S = \int_{t_0}^{t} dt(\frac{p \bullet p}{2m} - V_{(q)} - V_{qu}) \tag{8}$$

and

$$n_{(q,t)} = |\Psi|^2_{(q,t)} \tag{9}$$

can be derived [3,18] by the system of two coupled differential equations

$$\partial_t S_{(q,t)} = -V_{(q)} + \frac{\hbar^2}{2m} \frac{\nabla^2 |\Psi|_{(q,t)}}{|\Psi|_{(q,t)}} - \frac{1}{2m}(\nabla S_{(q,t)})^2 \qquad (10)$$

$$\partial_t |\Psi|_{(q,t)} = -\frac{1}{m} \nabla |\Psi|_{(q,t)} \cdot \nabla S_{(q,t)} - \frac{1}{2m} |\Psi| \nabla^2 S_{(q,t)} \qquad (11)$$

by taking the gradient of (10) and multiplying by $|\Psi|$ equation (11). It is straightforward to see that the system of equations (10-11) for the complex variable

$$\Psi_{(q,t)} = |\Psi|_{(q,t)} \exp[\frac{i}{\hbar} S_{(q,t)}] \qquad (12)$$

is equivalent to equal to zero the real and imaginary part of the Schrödinger equation

$$i\hbar \frac{\partial \Psi}{\partial t} = -\frac{\hbar^2}{2m} \nabla^2 \Psi + V_{(q)} \Psi. \qquad (13)$$

It is straightforward to see that (10) derives by the Schrödinger equation by using the following identity

$$\Psi^* \frac{\partial \Psi}{\partial t} + \Psi \frac{\partial \Psi^*}{\partial t} = -\frac{\hbar}{2im}(\Psi^* \nabla^2 \Psi - \Psi \nabla^2 \Psi^*) = -\frac{\hbar}{2im} \nabla \cdot (\Psi^* \nabla \Psi - \Psi \nabla \Psi^*)$$
$$= -\frac{\hbar}{2im} \nabla \cdot (\Psi^* \Psi \nabla \ln(\Psi / \Psi^*)) \qquad (15)$$

from where it follows that

$$\frac{\partial |\Psi|^2}{\partial t} = -\frac{1}{m} \nabla \cdot (|\Psi|^2 \nabla S) \qquad (16)$$

that by (7, 9) leads to (1).

In a similar way, except for null volume ensembles where $\Psi = 0$ and $\Psi^* = 0$, the following identity

$$\frac{1}{\Psi} \frac{\partial \Psi}{\partial t} - \frac{1}{\Psi^*} \frac{\partial \Psi^*}{\partial t} = -\frac{\hbar}{2im}(\frac{1}{\Psi} \nabla^2 \Psi + \frac{1}{\Psi^*} \nabla^2 \Psi^*) + \frac{2}{i\hbar} V_{(q)} = -\frac{2i}{\hbar} \frac{\partial S}{\partial t} \qquad (17)$$

leads to

$$\frac{\partial S}{\partial t} = \frac{\hbar^2}{4m}(\frac{1}{\Psi} \nabla^2 \Psi + \frac{1}{\Psi^*} \nabla^2 \Psi^*) - V_{(q)}$$
$$= \frac{\hbar^2}{4m}(\frac{1}{\Psi} \nabla^2 |\Psi| \exp(iS/\hbar) + \frac{1}{\Psi^*} \nabla^2 |\Psi| \exp(-iS/\hbar)) - V_{(q)} \qquad (18)$$
$$= -V_{(q)} + \frac{\hbar^2}{2m}(\frac{\nabla^2 |\Psi|}{|\Psi|} - \frac{\nabla S \cdot \nabla S}{\hbar^2})$$

that by (7, 9) leads to (10).

## 2.1. The hydrodynamic K-G equation for a free particle

In relativistic mechanics it is well known that the K-G equation reads

$$(\frac{1}{c}\frac{\partial}{\partial t})^2 \Psi - \nabla^2 \Psi = -\frac{m^2 c^2}{\hbar^2}\Psi \qquad (19)$$

more synthetically written as

$$\partial_{\sim}\partial^{\sim}\Psi = -\frac{m^2 c^2}{\hbar^2}\Psi \qquad (20)$$

where $\partial_{\sim} = (\frac{1}{c}\frac{\partial}{\partial t}, i\nabla)$.

In the same way as for (16), the current conservation equation can be found by the following identity

$$\Psi^*\partial_{\sim}\partial^{\sim}\Psi - \Psi\partial_{\sim}\partial^{\sim}\Psi^* = \partial_{\sim}(\Psi^*\partial^{\sim}\Psi - \Psi\partial^{\sim}\Psi^*) = \partial_{\sim}J^{\sim} = 0 \qquad (21)$$

where

$$J^{\sim} \propto \left(\frac{1}{c}(\Psi^*\frac{\partial}{\partial t}\Psi - \Psi\frac{\partial}{\partial t}\Psi^*), i(\Psi^*\nabla\Psi - \Psi\nabla\Psi^*)\right) = (J^0, iJ) \qquad (22)$$

that assuming $\propto = \frac{i\hbar}{2m}$ and

$$\ldots = \frac{J^0}{c} = \frac{i\hbar}{2mc^2}(\Psi^*\frac{\partial}{\partial t}\Psi - \Psi\frac{\partial}{\partial t}\Psi^*) = -\frac{|\Psi|^2}{mc^2}\frac{\partial S}{\partial t} \qquad (24.a)$$

$$J = \frac{\hbar}{2mi}(\Psi^*\nabla\Psi - \Psi\nabla\Psi^*) = \frac{\hbar}{2mi}\left(\Psi^*\!\!\mathcal{E}\nabla ln\frac{\Psi}{\Psi^*}\right) = |\Psi|^2 \frac{\nabla S}{m} = -\ldots\dot{q} \qquad (24.b)$$

leads to (see appendix A)

$$\frac{\partial\ldots}{\partial t} + \nabla\bullet(|\Psi|^2 \frac{\nabla S}{m}) \\ = \frac{\partial\ldots}{\partial t} + \nabla\bullet(\ldots\dot{q}) = 0 \qquad (25)$$

from where by (24) it follows that

$$p = \nabla S = \frac{1}{c^2}\frac{\partial S}{\partial t}\dot{q} \ . \qquad (26)$$

where for pure matter or antimatter (lower minus sign) states it holds

$$\frac{\partial S}{\partial t} = \pm E = E = \pm m\mathsf{x}c^2\sqrt{1 - \frac{V_{qu}}{mc^2}} \ . \qquad (27)$$

where

$$V_{qu} = \frac{\hbar^2}{m} \frac{\partial_\mu \partial^\mu |\Psi|}{|\Psi|}, \qquad (28)$$

Since the squared form of the operator of the K-G equation the Hamilton-Jacobi-like relations (2,3) cannot be maintained. Therefore, the K-G HQE does not own a classical-like form. Nevertheless, it remains as an alternative description with two real variables ($|\Psi|$ and $S$) [19] instead of a complex one $\Psi$ as in the K-G equation.

In order to end with the $|\Psi|$ and $S$ description, from (25) with the help of (23) that reads we obtain

$$\frac{\partial \left(|\Psi|^2 \frac{\partial S}{\partial t}\right)}{\partial t} + c^2 \nabla \cdot (|\Psi|^2 \nabla S) = 0 \qquad (29)$$

$$\frac{\partial \left(|\Psi|^2 \frac{\partial S}{\partial t}\right)}{\partial t} + \nabla \cdot (|\Psi|^2 \frac{\partial S}{\partial t} \dot{q}) = 0. \qquad (30)$$

As far as it concerns the action equation, we obtain (see appendix B)

$$\frac{1}{c^2} \frac{\partial^2 S}{\partial t^2} + \frac{1}{\hbar} \left(\frac{1}{c} \frac{\partial S}{\partial t}\right)^2 - \left(\frac{\nabla \cdot (|\Psi| \nabla S)}{|\Psi|}\right) + 2\hbar \left(\frac{1}{|\Psi|} \frac{1}{c} \frac{\partial |\Psi|}{\partial t}\right)^2 = 0 \qquad (31)$$

## 2.2. The hydrodynamic K-G equation for a charged particle

In order to obtain the hydrodynamic formulation of the K-G equation for a charged particle (zero spin) we proceed with the standard substitutions

$$H \to H - e\mathsf{W} \qquad (32)$$

$$p \to p - eA \qquad (33)$$

$$i\hbar \frac{\partial}{\partial t} \to i\hbar \frac{\partial}{\partial t} - e\mathsf{W} \qquad (34)$$

$$\frac{\hbar \nabla}{i} \to \frac{\hbar \nabla}{i} - eA \qquad (35)$$

and

$$\partial_\mu = (\frac{1}{c} \frac{\partial}{\partial t}, i\nabla) \to (\frac{1}{c}(\frac{\partial}{\partial t} + \frac{ie}{\hbar}\mathsf{W}), i(\nabla - \frac{ie}{\hbar}A) = \partial_\mu + \frac{ie}{\hbar} A_\mu = D_\mu \qquad (36)$$

so that the K-G equation for charged particle reads

$$D_\sim D^\sim \Psi = -\frac{m^2 c^2}{\hbar^2}\Psi. \tag{37}$$

It is straightforward to see that the relation

$$\Psi^* D_\sim D^\sim \Psi - \Psi D_\sim D^\sim \Psi^* = 0 \tag{38}$$

that with $J^\sim$ defined as

$$\begin{aligned}J^\sim &= \frac{i\hbar}{2m}\Psi^* D^\sim \Psi - \Psi(D^\sim \Psi)^* \\ &= \frac{i\hbar}{2m}\left(\frac{1}{c}(\Psi^*(\frac{\partial}{\partial t}+\frac{ie}{\hbar}W)\Psi - \Psi(\frac{\partial}{\partial t}-\frac{ie}{\hbar}W)\Psi^*), i(\Psi^*(\nabla-\frac{ie}{\hbar}A)\Psi - \Psi(\nabla+\frac{ie}{\hbar}A)\Psi^*)\right)\end{aligned} \tag{39}$$

leads to current conservation equation

$$\begin{aligned}\partial_\sim J^\sim &= \partial_\sim \left(\Psi^* \partial_\sim \partial^\sim \Psi - \Psi \partial_\sim \partial^\sim \Psi^*\right) - \frac{4ie}{\hbar}\Psi^* \partial_\sim A^\sim \Psi \\ &= \partial_\sim \left(\Psi^* \partial_\sim \partial^\sim \Psi - \Psi \partial_\sim \partial^\sim \Psi^*\right) - \frac{4ie}{\hbar}\Psi^*\left(\frac{1}{c^2}\frac{\partial W}{\partial t}+\nabla\cdot A\right)\Psi = 0\end{aligned} \tag{40}$$

and, hence, to

$$\frac{\partial\left(|\Psi|^2 \frac{\partial S}{\partial t}\right)}{\partial t} + c^2 \nabla \cdot (|\Psi|^2 \nabla S) = \frac{4ie}{\hbar}\left(\frac{1}{c^2}\frac{\partial W}{\partial t}+\nabla\cdot A\right)|\Psi|^2 \tag{41}$$

$$\frac{\partial\left(|\Psi|^2 \left(\frac{\partial S}{\partial t}+eW\right)\right)}{\partial t} + c^2 \nabla \cdot (|\Psi|^2 (\nabla S - eA)) = 0. \tag{42}$$

Moreover, by applying the rules (32-35) to $S = -\frac{i\hbar}{2}ln\frac{\Psi}{\Psi^*}$ and to $|\Psi| = (\Psi\Psi^*)^{\frac{1}{2}}$ and recovering the rules

$$\frac{\partial |\Psi|^2}{\partial t} \rightarrow \frac{\partial |\Psi|^2}{\partial t} \tag{43}$$

$$\frac{\partial S}{\partial t} \rightarrow \frac{\partial S}{\partial t} + eW \tag{44}$$

$$\nabla S \rightarrow \nabla S - eA \tag{45}$$

$$\frac{\partial^2 S}{\partial t^2} \rightarrow \frac{\partial\left(\frac{\partial S}{\partial t} + e\mathrm{W}\right)}{\partial t}, \tag{46}$$

from (30) it follows that

$$\frac{1}{c^2}\frac{\partial\left(\frac{\partial S}{\partial t} + e\mathrm{W}\right)}{\partial t} - \frac{1}{\hbar c^2}\left(\frac{\partial S}{\partial t} + e\mathrm{W}\right)^2 - \left(\nabla\bullet(\nabla S - eA) + (\nabla S - eA)\bullet\frac{\nabla|\Psi|}{|\Psi|}\right) + 2\hbar\left(\frac{1}{|\Psi|}\frac{1}{c}\frac{\partial|\Psi|}{\partial t}\right)^2 = 0 \tag{47}$$

that with equation (42) represent the hydrodynamic K-G equations for a charged particle.

## 3. The classical limit

In order to derive the classical limit of the K-G hydrodynamic equations, we will use the following limiting expressions

$$\lim_{\frac{q}{c}\to 0} \frac{\partial S}{\partial t} = \pm \lim_{\frac{q}{c}\to 0} mc^2 \mathsf{x} = \pm mc^2 \tag{48}$$

$$\lim_{\frac{q}{c}\to 0} \ldots = -\lim_{\frac{q}{c}\to 0} \frac{|\Psi|^2}{mc^2}\frac{\partial S}{\partial t} = -\lim_{\frac{q}{c}\to 0} \pm \mathsf{x}\,|\Psi|^2 = \pm|\Psi|^2 \tag{49}$$

with $\mathsf{x} = \dfrac{1}{\sqrt{1 - \dfrac{\dot{q}^2}{c^2}}}$ , from where we obtain

$$\frac{\partial\left(|\Psi|^2\,mc^2\right)}{\partial t} + c^2\nabla\bullet(|\Psi|^2\,\nabla S) = 0 \tag{50}$$

and, hence,

$$\frac{\partial|\Psi|^2}{\partial t} + \nabla\bullet(|\Psi|^2\,\frac{\nabla S}{m}) = 0 \tag{51}$$

that taking into account for the mass phase factor of the K-G wave function $\Psi$ respect to the Schrödinger one Œ that reads

$$\Psi\pm = exp[-\frac{\pm imc^2 t}{\hbar}]\text{Œ}_\pm \tag{52}$$

so that $|\Psi|=|\text{Œ}|$, it follows that

$$\frac{\partial|\text{Œ}|^2}{\partial t} + \nabla\bullet(|\text{Œ}|^2\,\frac{\nabla S}{m}) = 0, \tag{53}$$

As far as it concerns the action equation,

$$\frac{1}{c^2}\frac{\partial^2 S}{\partial t^2} + \frac{1}{\hbar}\left(\frac{1}{c}\frac{\partial S}{\partial t}\right)^2 - 2\hbar\left(\frac{1}{|\Psi|}\frac{1}{c}\frac{\partial |\Psi|}{\partial t}\right)^2 - \frac{\nabla^2\Psi}{\Psi} - \frac{\nabla^2\Psi^*}{\Psi^*} = 0 \qquad (54)$$

with the classical approximation of the K-G equation (i.e., $\frac{\nabla^2}{\Psi}\Psi \cong 0$) we obtain that

$$\frac{\partial^2 S}{\partial t^2} - \frac{1}{\hbar}\left(\frac{\partial S}{\partial t}\right)^2 + 2\hbar\left(\frac{1}{|\Psi|}\frac{\partial |\Psi|}{\partial t}\right)^2 = \frac{c^2 \nabla^2 \Psi}{\Psi} - \frac{c^2 \nabla^2 \Psi^*}{\Psi^*} \cong 0 \qquad (55)$$

and, given that

$$2\hbar\left(\frac{1}{|\Psi|}\frac{\partial |\Psi|}{\partial t}\right) = \hbar\left(\frac{1}{\Psi}\frac{\partial \Psi}{\partial t} + \frac{1}{\Psi^*}\frac{\partial \Psi^*}{\partial t}\right) = \hbar\left(\frac{1}{\Psi}\frac{\partial \Psi}{\partial t} - i\frac{mc^2}{\hbar} + \frac{1}{\Psi^*}\frac{\partial \Psi^*}{\partial t} + i\frac{mc^2}{\hbar}\right) \qquad (56)$$

that with the classical relation (see equation C.4 in appendix C)

$$\left((\frac{1}{c}\frac{\partial}{\partial t}) - i\frac{mc}{\hbar} + i\hbar\frac{\nabla^2}{2mc}\right)\Psi = 0 \qquad (57)$$

$$\left((\frac{1}{c}\frac{\partial}{\partial t}) + i\frac{mc}{\hbar} - i\hbar\frac{\nabla^2}{2mc}\right)\Psi^* = 0 \qquad (58)$$

leads to

$$\left(\frac{1}{|\Psi|}\frac{\partial |\Psi|}{\partial t}\right) = \frac{1}{2}\left(\frac{1}{\Psi}\frac{\partial \Psi}{\partial t} + \frac{1}{\Psi^*}\frac{\partial \Psi^*}{\partial t}\right) = \frac{i\hbar}{2}\left(\frac{1}{\Psi}\frac{\nabla^2}{2m}\Psi - \frac{1}{\Psi^*}\frac{\nabla^2}{2m}\Psi^*\right), \qquad (60)$$

and since

$$\frac{\partial^2 S}{\partial t^2} \cong 0 \qquad (61)$$

it follows that

$$\left(\frac{\partial S}{\partial t}\right)^2 = 2\hbar^2\left(\frac{1}{|\Psi|}\frac{1}{c}\frac{\partial |\Psi|}{\partial t}\right)^2 = -\frac{\hbar^4}{4}\left(\frac{1}{\Psi}\frac{\nabla^2}{2m}\Psi - \frac{1}{\Psi^*}\frac{\nabla^2}{2m}\Psi^*\right)^2 \qquad (62)$$

that with the classical K-G approximation (i.e., $\frac{\nabla^2}{\Psi^*}\Psi^* \cong 0$), finally gives

$$\left(\frac{\partial S}{\partial t}\right) = -\frac{\hbar^2}{4m}\left(\frac{\nabla^2}{\Psi}\Psi + \frac{\nabla^2}{\Psi*}\Psi* - 2\frac{\nabla^2}{\Psi*}\Psi*\right)$$

$$\cong -\frac{\hbar^2}{4m}\left(\frac{\nabla^2}{\Psi}\Psi + \frac{\nabla^2}{\Psi*}\Psi*\right) = \frac{i\hbar}{2}\left(\frac{1}{\Psi}\frac{\partial \Psi}{\partial t} - \frac{1}{\Psi*}\frac{\partial \Psi*}{\partial t}\right) \qquad (63)$$

The equivalent expression for the Schrödinger wave function is obtained by taking into account for the mass phase factor that leads to

$$\frac{\partial S}{\partial t} \cong \frac{i\hbar}{2}\frac{\partial ln[\frac{\Psi}{\Psi*}]}{\partial t} = \frac{i\hbar}{2}\frac{\partial\left(ln[\frac{Œ}{Œ*}] - \frac{2imc^2}{\hbar}t\right)}{\partial t} = mc^2 + \frac{i\hbar}{2}\frac{\partial ln[\frac{Œ}{Œ*}]}{\partial t} = mc^2 - \frac{\partial S_{cl}}{\partial t} \quad (64)$$

where $S_{cl} = \frac{p^2}{2m}t$. From (64) we obtain

$$\frac{\partial S_{cl}}{\partial t} = \frac{i\hbar}{2}\frac{\partial ln[\frac{Œ}{Œ*}]}{\partial t} = \frac{i\hbar}{2}\left(\frac{1}{Œ}\frac{\partial Œ}{\partial t} - \frac{1}{Œ*}\frac{\partial Œ*}{\partial t}\right) = \frac{1}{2}\left(\frac{1}{Œ}HŒ - \frac{1}{Œ*}H^+Œ*\right) (65)$$

that by using the identity for the Hamiltonian operator

$$H = \frac{\partial S_{cl}}{\partial t} \cong \frac{p^2}{2m} \rightarrow -\frac{\hbar^2\nabla^2}{2m} \qquad (66)$$

leads to

$$\frac{\partial S_{cl}}{\partial t} = \frac{\hbar^2}{4m}\left(\frac{1}{Œ}\nabla^2 Œ - \frac{1}{Œ*}\nabla^2 Œ*\right) = \frac{\hbar^2}{2m}\left(\frac{\nabla^2 |Œ|}{|Œ|} - \frac{\nabla S \cdot \nabla S}{\hbar^2}\right). \qquad (67)$$

That is the classical hydrodynamic equation (10).
For a charged particle, by using the substitutions

$$\frac{\partial |Œ|^2}{\partial t} \rightarrow \frac{\partial |Œ|^2}{\partial t} \qquad (68)$$

$$\frac{\partial S}{\partial t} \rightarrow \frac{\partial S}{\partial t} + eW \qquad (69)$$

$$\nabla S \rightarrow \nabla S - eA \qquad (70)$$

we end with

$$\frac{\partial S_{cl}}{\partial t} + eW + \frac{\hbar^2}{2m}\frac{(\nabla S - eA)^2}{\hbar^2} = \frac{\hbar^2}{2m}\frac{\nabla^2 |Œ|}{|Œ|} \qquad (71)$$

that represents the correct classical limit [19].

## 4. Discussion

If we look at the manageability of the quantum equations no one would solve the hydrodynamic ones. Nevertheless, the interest for the QHA remained unaltered along the time. The motivation for this does not only reside in the formal analogy with the classical mechanics, but also in the fact that the non local properties of quantum mechanics are more clearly mathematically recognizable in the model.
In the classical limit, in order to establish the hydrodynamic "mechanical" analogy, the gradient of (10) is taken. When we do that, we broaden the solutions of (10) so that not every solution of the hydrodynamic equations can be a solution of the Schrödinger problem.

In fact, the state of a particle in the QHA is defined by the four real functions n and $\dot{q} = \frac{\partial H}{\partial p} = \frac{\nabla S_{(q,t)}}{m}$.

The restriction of the class of solutions of the classical hydrodynamic analogy comes from additional conditions such as the quantization condition (non local one) on the action. The integrability of the action gradient in order to warrant the existence of the scalar action function is warranted if the probability fluid is irrotational, that being

$$\dot{q} = \frac{\nabla S_{(q,t)}}{m} \qquad (72)$$

and

$$S_{(q,t)} = m \int_{q_0}^{q} dl \cdot \dot{q} \qquad (73)$$

it is to say that

$$\nabla \times \dot{q} = 0 \qquad (74)$$

and hence that

$$\Gamma c = m \oint dl \cdot \dot{q} = 0 \qquad (75)$$

Moreover, since the action is contained in the argument of the exponential function of the wave function, all the multiples of $2f\hbar$, with $\Gamma c = 0 \pm 2f\hbar n$ on a closed contour, are accepted.
In the QHA, these non local characteristics of dynamics are transferred to the dynamics through the quantum the potential (6) to which the quantized action is linked.
In the Schrödinger problem not all solutions are considered but only those that fulfill precise boundary conditions (e.g., for the harmonic oscillators the eigenstates are those that goes to zero to infinity).
In the QHA the eigenstates are defined by the stationarity that happens when the force generated by the quantum potential exactly counterbalance that one due to the Hamiltonian potential. Since the quantum potential changes with the state of the system, more the one stationary state is possible (and more than one quantized values of the action may exist).
In the QHA the non locality does not come from boundary ones (that are apart from the equations) but from the quantum pseudo-potential (6) that depends by the state of the system and is a source of an elastic-like energy for it [20]. If we consider a bi-dimensional space, the quantum potential makes the vacuum acting like an elastic membrane that becomes quite brittle on very small scale.
Since the force of the quantum potential in a point depends by the state of the system around it, the character of non-local dynamics is introduced into the equations.
In force of this, in the QHA the non local properties can be very well identified (and hence studied) in precise mathematic terms.
As far as it concerns the relativistic limit, some observations must be done.
Due to the squared operational form of the K-G equation, the Hamilton-Jacobi structure (like relations (2,3)) cannot be retrieved (even with additional non-local terms) in the relativistic limit. This is due to the presence of anti-particles with negative energy that contribute to the superposition of quantum states.

Nevertheless, it remains unaltered the possibility of having an alternative description with two real variables (/Œ / and $S$) [19] instead of a complex one with $\Psi$ as in the K-G equation .

The widening of the solutions by defining an equation for the gradient of the action, like in the classical approach, does not find other needs that in ending with equations owing a formal classic-like structure. Actually, we do not need to do it at all. The description with two real variables /Œ / and $S$ has the advantage of putting in evidence the "source" terms leading to the non-local dynamics

$$\frac{\nabla /\Psi/}{/\Psi/} \tag{76}$$

$$\frac{1}{/\Psi/}\frac{1}{c}\frac{\partial /\Psi/}{\partial t} \tag{77}$$

in equation (47) and of having the action defined without the need to be derived by integration of its gradient. Due to this, the relativistic hydrodynamic description can be very useful in studying the non-local properties of relativistic quantum mechanics. This fact becomes more evident in presence of fluctuations.

If in the case of null fluctuations the QHA approach clearly shows that exists an ensemble of forbidden values for the action (see figure 1).

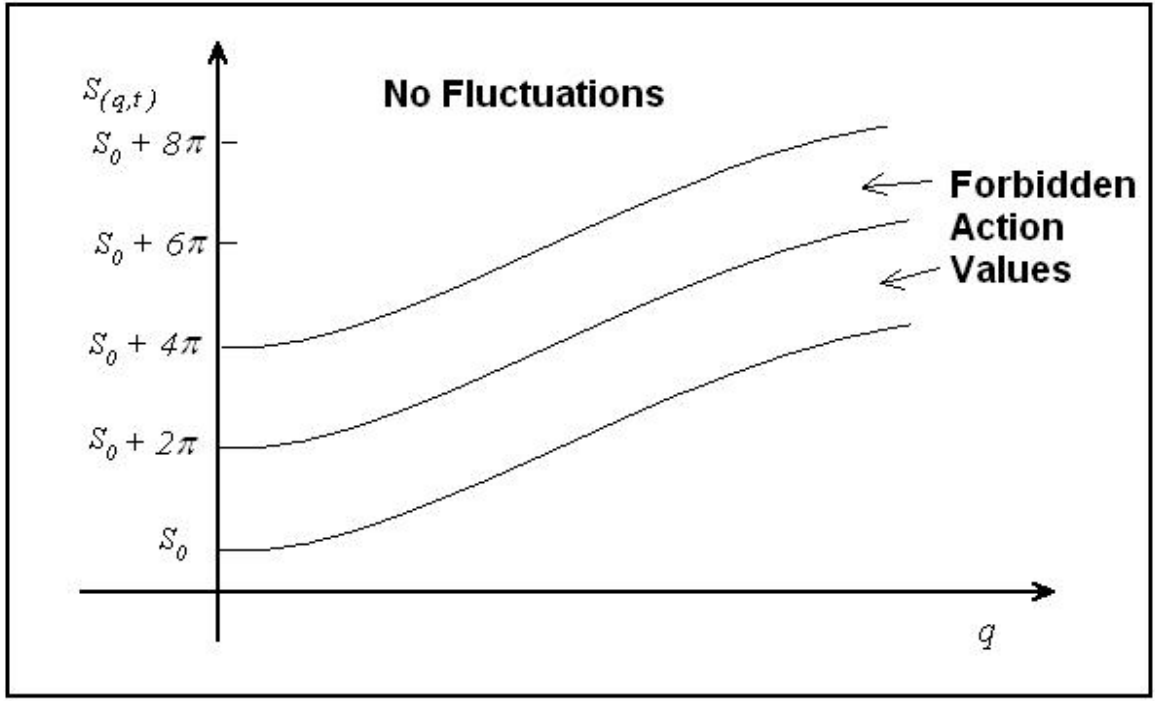

Figure 1. Quantized action in the case of a system without fluctuations

In the stochastic generalization of the hydrodynamic model [11] (where the action of the system undergoes to fluctuation (see figure 2))

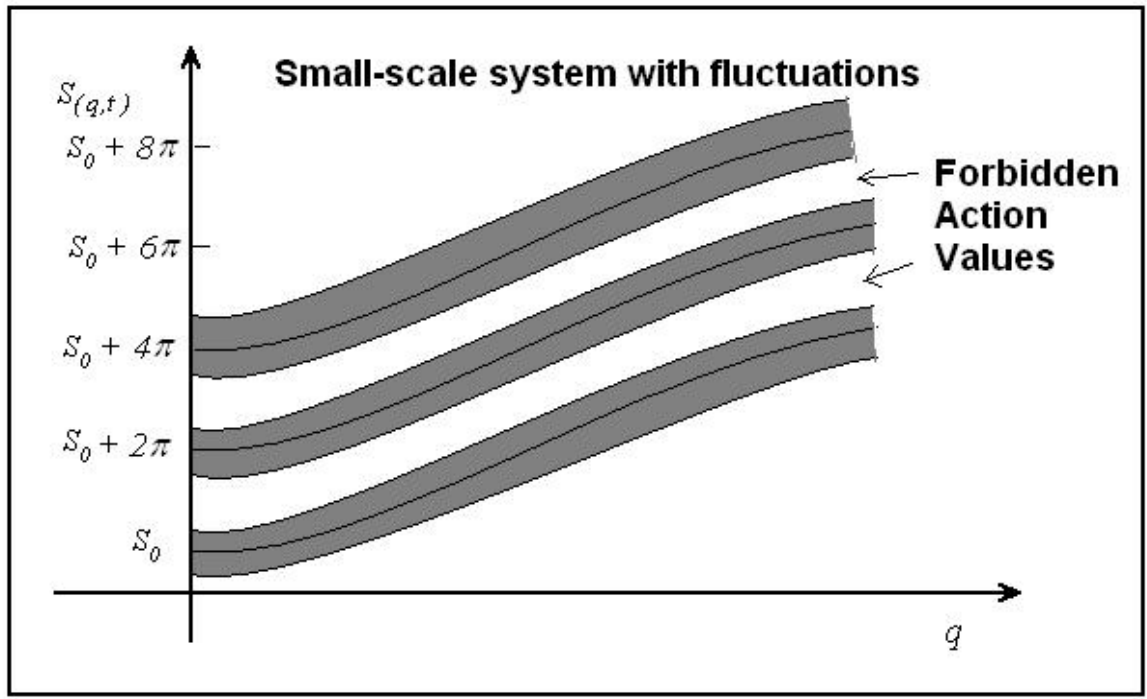

Figure 2. Quantized action in the case of a small-scale system submitted to fluctuations

depending on the fluctuations amplitude and energy eigenvalues gaps, the "classical freedom" can be achieved on large scale system [11] (see figure 3).

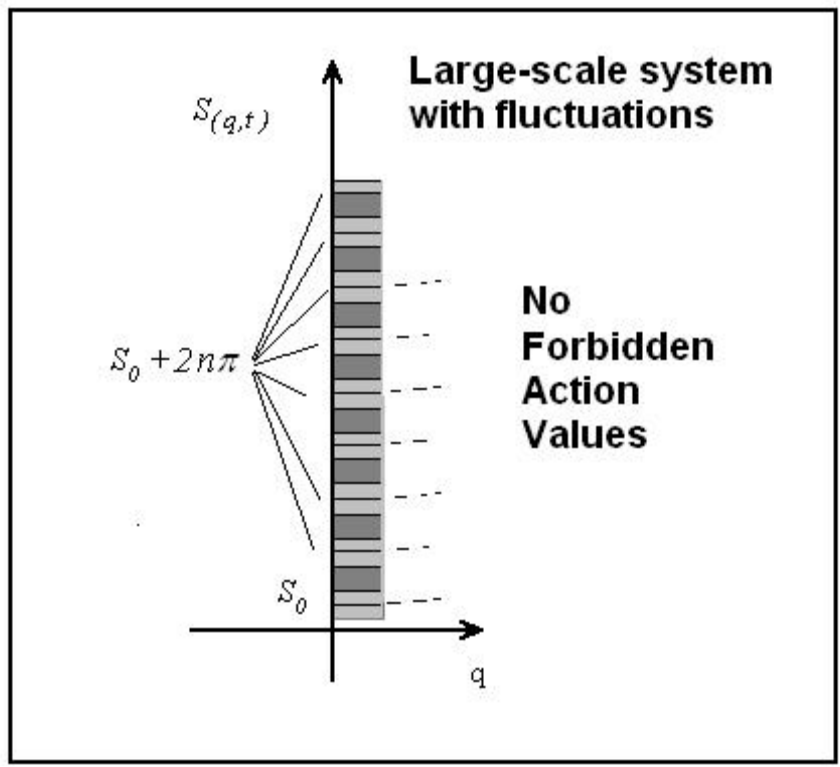

Figure 3. Quantized action in the case of a large-scale system submitted to fluctuations

The transition to the "deterministic" limit given in figure1, passing through the configuration shown in figure 2, is generated by the property of quantum potential that makes the vacuum poorly "flexible" on small scale (i.e., Compton length) [11] so that the noise is hindered and suppressed and the quantization condition recovers its effectiveness.

The above behavior has been analytically investigated in the classical QHA [11]. Its extension to the relativistic limit can be useful to fully understand the co-existence of the quantum and classical dynamics. It is worth mentioning that the quantum to classical transition, with the breaking of the quantum entanglement, in the relativistic limit is logically deputed to facilitate the understanding of the EPR paradox.

# 6. Conclusion

In the present paper they have been derived the two coupled hydrodynamic-type quantum equations for the phase and the amplitude of the wave function, of the relativistic Klein-Gordon equation.

The work shows that in classical limit the Madelung pseudo-potential [1] as well as the quantum pseudo-potential for a charged particle [2,3] are recovered.

The description of the non-local interactions of quantum mechanics in the hydrodynamic model is discussed both for free and charged particles. The implications to the stochastic quantum case are preliminary depicted.

## Nomenclature

| | |
|---|---|
| n = square wave function modulus | number of particle $l^{-3}$ |
| $S$ = action of the system | $m^{-1} l^{-2} t$ |
| $m$ = mass particle | m |
| $\hbar$ = Plank's constant | $m\, l^2\, t^{-1}$ |
| c = light speed | $l\, t^{-1}$ |
| $H$ = Hamiltonian of the system | $m\, l^2\, t^{-2}$ |
| $V$ = potential energy | $m\, l^2\, t^{-2}$ |
| $V_{qu}$ = quantum potential energy | $m\, l^2\, t^{-2}$ |

## References

1. Madelung, E.: Quanten theorie in hydrodynamische form (Quantum theory in the hydrodynamic form). Z. Phys. 40, 322-6 (1926).

2. Jánossy, L.: Zum hydrodynamischen Modell der Quantenmechanik. Z. Phys. 169, 79 (1962).8. Gardner, C.L.: The quantum hydrodynamic model for semiconductor devices. SIAM J. Appl. Math. 54, 409 (1994).

3. I. Bialyniki-Birula, M., Cieplak, J., Kaminski, "Theory of Quanta", Oxford University press, Ny, (1992), p. 87-111.

4. Tsekov, R., Bohmian Mechanics Versus Madelung Quantum Hydrodynamics, arXiv:0904.0723v8 [quantum-phys] (2011).

5. Bousquet D, Hughes KH, Micha DA Burghardt I. Extended hydrodynamic approach to quantum-classical nonequilibrium evolution I. Theory. J. Chem. Phys. 2001;134.

6. Wyatt RE. Quantum dynamics with trajectories: Introduction to quantum hydrodynamics, Springer, Heidelberg; 2005.

7. Gardner CL. The quantum hydrodynamic model for semiconductor devices. SIAM J. Appl. Math. 1994;54:409.

8. Bertoluzza, S. and Pietra, P.: Space-Frequency Adaptive Approximation for Quantum Hydrodynamic Models. Reports of Institute of Mathematical Analysis del CNR, Pavia, Italy, (1998).

9. Morato LM, Ugolini S, Stochastic description of a Bose–Einstein Condensate, Annales Henri Poincaré. 2011;12(8):1601-1612.

10. Tamura, H., Ramon, J. G. S., Bittner, E., Bourghardt, I., Phys Rev. Lett. 100,107402 (2008).

11. Chiarelli, P., "Can fluctuating quantum states acquire the classical behavior on large scale?" J. Adv. Phys. 2013; **2**, 139-163 .

12. A. Mariano, P. Facchi, and S. Pascazio Decoherence and Fluctuations in Quantum Interference Experiments, Fortschr. Phys. 49 (2001) 10—11, 1033 — 1039

13. M. Brune, E. Hagley, J. Dreyer, X. Maıˆtre, A. Maali, C. Wunderlich, J. M. Raimond, and S. Haroche Observing the Progressive Decoherence of the "Meter" in a Quantum Measurement, Phys Rev Lett **77** 24 ( 1996)

14. E. Calzetta and B. L. Hu, Quantum Fluctuations, Decoherence of the Mean Field, and Structure Formation in the Early Universe, Phys.Rev.D, **52**, 6770-6788, (1995).

15. C., Wang, P., Bonifacio, R., Bingham, J., T., Mendonca, Detection of quantum decoherence due to spacetime fluctuations, 37[th] COSPAR Scientific Assembly. Held 13-20 July 2008, in Montréal, Canada., p.3390.

16. F., C., Lombardo , P. I. Villar, Decoherence induced by zero-point fluctuations in quantum Brownian motion, Physics Letters A 336 (2005) 16–24

17. Cerruti, N.R., Lakshminarayan, A., Lefebvre, T.H., Tomsovic, S.: Exploring phase space localization of chaotic eigenstates via parametric variation. Phys. Rev. E 63, 016208 (2000).

18. Weiner, J.H., *Statistical Mechanics of Elasticity* (John Wiley & Sons, New York, 1983), p. 315-317.

19. Ibid [3] pp. 90

20. Weiner, J.H., *Statistical Mechanics of Elasticity* (John Wiley & Sons, New York, 1983), p. 406-411.

# Appendix A

By inspection on the solutions of K-G equation for the free particle,

$$\Psi = |\Psi| exp[-\frac{iEt}{\hbar}] exp[-\frac{ip \cdot x}{\hbar}] \qquad (A.1)$$

with $V_{qu} = \frac{\hbar^2}{m}\frac{\partial_\mu \partial^\mu |\Psi|}{|\Psi|} = 0$, considering also the negative energy values

$$E = \pm mc^2 \qquad (A.2)$$

that leads to

$$-\frac{\partial S}{\partial t} = E = \pm mc^2 , \qquad (A.3)$$

we have

$$\ldots = \frac{J^0}{c} = \pm \chi \, |\Psi|^2 \tag{A.4}$$

so that, by defining $S_{\pm} = \pm S$, we can write

$$\frac{\partial \ldots}{\partial t} - \frac{i\hbar}{2m} \nabla \cdot (\Psi^* \Psi \nabla \ln(\Psi/\Psi^*)) = \frac{\partial (\pm \chi \, |\Psi|^2)}{\partial t} + \frac{1}{m} \nabla \cdot (|\Psi|^2 \, \nabla S_{\pm})$$
$$= \frac{\partial (\pm \chi \, |\Psi|^2)}{\partial t} + \nabla \cdot (\pm \chi \, |\Psi|^2 \, \frac{\pm \nabla S}{\pm m\chi}) = \pm \frac{\partial (\chi \, |\Psi|^2)}{\partial t} \pm \nabla \cdot (\chi \, |\Psi|^2 \, \frac{\nabla S}{m\chi}) \tag{A.5}$$

that univocally reads

$$\frac{\partial \ldots}{\partial t} + \nabla \cdot (\ldots \dot{q}) = 0 \tag{A.6}$$

# Appendix B

By using the identity

$$\frac{\partial \left( \frac{1}{\Psi} \frac{\partial \Psi}{\partial t} - \frac{1}{\Psi^*} \frac{\partial \Psi^*}{\partial t} \right)}{\partial t} + \left( \frac{1}{\Psi} \frac{\partial \Psi}{\partial t} - \frac{i}{\Psi^*} \frac{\partial \Psi^*}{\partial t} \right)^2 + 2i \frac{1}{\Psi} \frac{\partial \Psi}{\partial t} \frac{1}{\Psi^*} \frac{\partial \Psi^*}{\partial t} = + \frac{1}{\Psi} \frac{\partial^2 \Psi}{\partial t^2} - \frac{1}{\Psi^*} \frac{\partial^2 \Psi^*}{\partial t^2}$$

$$= \frac{2i}{\hbar} \frac{\partial^2 S}{\partial t^2} + \frac{(i-1)^2}{\hbar^2} \left( \frac{\partial S}{\partial t} \right)^2 + 2i \frac{\partial \left( \ln|\Psi| + \frac{iS}{\hbar} \right) \partial \left( \ln|\Psi| - \frac{iS}{\hbar} \right)}{\partial t} \frac{}{\partial t} = + \frac{1}{\Psi} \frac{\partial^2 \Psi}{\partial t^2} - \frac{1}{\Psi^*} \frac{\partial^2 \Psi^*}{\partial t^2}$$

$$= \frac{2i}{\hbar} \frac{\partial^2 S}{\partial t^2} + \frac{(i-1)^2}{\hbar^2} \left( \frac{\partial S}{\partial t} \right)^2 + 4i \left( \left( \frac{1}{|\Psi|} \frac{\partial |\Psi|}{\partial t} \right)^2 + \frac{1}{\hbar^2} \left( \frac{\partial S}{\partial t} \right)^2 \right) = \frac{c^2 \nabla^2 \Psi}{\Psi} - \frac{c^2 \nabla^2 \Psi^*}{\Psi^*}$$
(B.1)

$$\frac{2i}{\hbar} \frac{1}{c^2} \frac{\partial^2 S}{\partial t^2} + \frac{(i-1)^2}{\hbar^2} \left( \frac{1}{c} \frac{\partial S}{\partial t} \right)^2 + 4i \left( \left( \frac{1}{|\Psi|} \frac{1}{c} \frac{\partial |\Psi|}{\partial t} \right)^2 + \frac{1}{\hbar^2} \left( \frac{1}{c} \frac{\partial S}{\partial t} \right)^2 \right)$$
$$= \frac{\nabla^2 \Psi}{\Psi} - \frac{\nabla^2 \Psi^*}{\Psi^*} = \frac{2i}{\hbar} \left( \nabla \cdot \nabla S + \nabla S \cdot \frac{\nabla |\Psi|}{|\Psi|} \right) = \frac{2i}{\hbar} \left( \frac{\nabla \cdot (|\Psi| \nabla S)}{|\Psi|} \right) \tag{B.2}$$

we can finally write

$$\frac{2i}{\hbar}\frac{1}{c^2}\frac{\partial^2 S}{\partial t^2} + \frac{(i+1)^2}{\hbar^2}\left(\frac{1}{c}\frac{\partial S}{\partial t}\right)^2 - \frac{2i}{\hbar}\left(\frac{\nabla\bullet(|\Psi|\nabla S)}{|\Psi|}\right) + 4i\left(\frac{1}{|\Psi|}\frac{1}{c}\frac{\partial|\Psi|}{\partial t}\right)^2 = 0 \qquad (B.3)$$

$$\frac{1}{c^2}\frac{\partial^2 S}{\partial t^2} + \frac{1}{\hbar}\left(\frac{1}{c}\frac{\partial S}{\partial t}\right)^2 - \left(\frac{\nabla\bullet(|\Psi|\nabla S)}{|\Psi|}\right) + 2\hbar\left(\frac{1}{|\Psi|}\frac{1}{c}\frac{\partial|\Psi|}{\partial t}\right)^2 = 0 \qquad (B.4)$$

# Appendix C

By applying the classical approximation to the K-G equation that reads

$$\left((\frac{1}{c}\frac{\partial}{\partial t}) + i\frac{mc}{\hbar}\right)\left((\frac{1}{c}\frac{\partial}{\partial t}) - i\frac{mc}{\hbar}\right)\Psi$$
$$= \left((\frac{1}{c}\frac{i}{\hbar}\left(\frac{\partial S_-}{\partial t}\right)_{op}) + i\frac{mc}{\hbar}\right)\left((\frac{1}{c}\frac{i}{\hbar}\left(\frac{\partial S_+}{\partial t}\right)_{op}) - i\frac{mc}{\hbar}\right)\Psi = \nabla^2\Psi \cong 0 \qquad (C.1)$$

(where the suffix "*op*" stands for operators and $S_\pm \cong \pm S$ ) and by introducing the classical limit of the relativistic action

$$S \cong \left(mc^2 + \frac{p^2}{2m}\right)t \to S_{op} \cong \left(mc^2 - \frac{\hbar^2}{2m}\nabla^2\right)t \qquad (C.2)$$

it is possible to write

$$\left((\frac{1}{c}\frac{\partial}{\partial t}) + i\frac{mc}{\hbar}\right)\left((\frac{1}{c}\frac{\partial}{\partial t}) - i\frac{mc}{\hbar}\right)\Psi$$
$$= \left((-\frac{i}{\hbar}\left(mc - \frac{\hbar^2}{2mc}\nabla^2\right)) + i\frac{mc}{\hbar}\right)\left((\frac{i}{\hbar}\left(mc - \frac{\hbar^2}{2mc}\nabla^2\right)) - i\frac{mc}{\hbar}\right)\Psi \qquad (C.3)$$
$$= \left((i\frac{\hbar}{2mc}\nabla^2)(-i\frac{\hbar}{2mc}\nabla^2)\right)\Psi$$

$$\left((\frac{1}{c}\frac{\partial}{\partial t}) + i\frac{mc}{\hbar} - i\frac{\hbar}{2mc}\nabla^2\right)\left((\frac{1}{c}\frac{\partial}{\partial t}) - i\frac{mc}{\hbar} + i\frac{\hbar}{2mc}\nabla^2\right)\Psi = 0 \qquad (C.4)$$

that at zero order reads

$$\left((\frac{1}{c}\frac{\partial}{\partial t}) + i\frac{mc}{\hbar}\right)\left((\frac{1}{c}\frac{\partial}{\partial t}) - i\frac{mc}{\hbar}\right)\Psi = \left(-i\frac{\hbar}{2mc}\nabla^2\right)\left(i\frac{\hbar}{2mc}\nabla^2\right)\Psi \cong 0. \qquad (C.5)$$